\documentclass{article}

\usepackage{PRIMEarxiv}

\usepackage[utf8]{inputenc} 
\usepackage[T1]{fontenc}    
\usepackage{hyperref}       
\usepackage{url}            
\usepackage{booktabs}       
\usepackage{amsfonts}       
\usepackage{nicefrac}       
\usepackage{microtype}      
\usepackage{lipsum}
\usepackage{fancyhdr}       
\usepackage{graphicx}       
\graphicspath{{media/}}     

\title{Analyzing DCTCP and Cubic Buffer Sharing under Diverse Router Configurations
}

\author{
  Santiago Vargas, Aruna Balasubramanian \\
  Stony Brook University \\
  \texttt{\{savargas, arunab\}@cs.stonybrook.edu} \\
   \And
  Srikanth Sundaresan \\
  Meta \\
  \texttt{ssundaresan@meta.com} \\
}

\begin{document}
\maketitle

\begin{abstract}
In this work, we look at the impact of router configurations on DCTCP and Cubic traffic when both algorithms share router buffers in the data center.
Modern data centers host traffic with mixed congestion controls, including DCTCP and Cubic traffic.
Both DCTCP and Cubic in the data center can compete with each other and potentially starve and/or be unfair to each other when sharing buffer space in the data center.
This happens since both algorithms are at odds with each other in terms of buffer utilization paradigms where DCTCP attempts to limit buffer utilization while Cubic generally fills buffers to obtain high throughput.
As a result, we propose methods for a measurement-driven analysis of DCTCP and Cubic performance when sharing buffers in data center routers via simulation.
We run around 10000 simulation experiments with unique router configurations and network conditions.
Afterwards, we present a generalizable ML model to capture the effect that different buffer settings have on DCTCP and Cubic streaming traffic in the data center. 
Finally, we suggest that this model can be used to tune buffer settings in the data center.
\end{abstract}


\section{Introduction}
Data centers are home to a diverse set of applications including data processing, storage, machine learning, computation, etc.
At there core, data center networks host the connectivity behind these applications with the goal of providing high bandwidth and low latency.

Towards the goal of providing low-latency data center networks, DCTCP has been largely adopted in the data center~\cite{alizadeh2010data, judd2015attaining}.
This relatively newer congestion control reacts faster to buffer congestion than loss-based algorithms due to the use of ECN packets.
These packets are able to inform TCP senders that network congestion is occurring, allowing them to throttle sending rates much before receiving a packet loss signal. 
As a result, ECN based algorithms like DCTCP enable network operators to tune the data center for lower-latency due to decreased congestion while also achieving high bandwidth for flows.

However, current data center networks do not solely contain DCTCP traffic and may have a mix of both ECN-based and loss-based algorithms, like Cubic, sharing data center networks~\cite{ghabashnehIMC22}.
The issue arises in that both of these algorithms have distinct control mechanisms, ECN and loss feedback, with largely divergent goals.
For DCTCP, ECN feedback is utilized to minimize buffer usage (instead of loss) in order to improve latency and sustain high bandwidth.
Conversely, loss-based algorithms send as much as possible until reaching a loss, which may cause larger buffer utilization in order to minimize losses and maximize sending rates.

These two distinct congestion feedback mechanisms present the following dichotomy:
On the one hand, ECN-based congestion controls attempt to minimize queuing congestion in network buffers by providing earlier feedback via ECN about buffer congestion.
On the other hand, loss-based algorithms work by saturating router buffers until packets are lost and then throttling sending.
Thus loss-based algorithms are seemingly at odds with ECN-based algorithms as they require certain buffer capacity to achieve target sending rates and avoid losses which cut their sending rate.
Therefore, network operators have the important task of balancing the buffer sharing behaviors of loss-based and ECN-based algorithm in data center network buffers.

In this work, we examine buffer sharing through the lens of DCTCP and Cubic since DCTCP is a popular, well-studied ECN-based algorithm and Cubic is the Linux default and most popular loss-based algorithm.
We seek to answer the following question: \emph{How can network operators improve the buffer sharing behavior of DCTCP and Cubic in the data center?}

In order to better understand the dynamic between DCTCP and Cubic buffer sharing in the data center, we first propose examining a wide range of data center router configurations, namely ECN and RED settings, which control DCTCP and Cubic behavior at the buffering level.
ECN controls the buffer threshold for marking packets as congested.
RED~\cite{floydRED}, which stands for Random Early Detection, attempts to provide loss feedback earlier than normal by randomly dropping packets before completely filling router buffers.
We specifically explore ECN and RED settings because these are well-studied queue management mechanisms commonly found in commodity data center switches.

In total, we simulate around 10000 buffer sharing experiments each with unique router buffer configurations and varying network conditions to have a holistic understanding of DCTCP and Cubic buffer sharing behavior.

The main results from these experiments is that an optimal ECN and RED settings for DCTCP and Cubic buffer sharing is not immediately obvious.
We find that there are configurations for complete DCTCP buffer domination and complete Cubic buffer saturation.
Likewise, these buffering configurations can significantly affect loss and hurt overall performance.

Next, our simulation experiments have large runtimes and overheads due to the large data center BDPs being used in the experiments.
Since these experiments take a significant time to run, it becomes difficult to run more fine-grained experiments and further explore the search space of router configurations.
Therefore, we propose using Machine Learning to model the buffering interaction between DCTCP and Cubic flows under different router configurations and network conditions.
An ML model can provide us with quick inference and be generalizable to extend to unexplored settings.

We phrase the learning problem as a regression where router configurations and network conditions are input features and buffer sharing performance, including share of Cubic vs. DCTCP throughput and buffer utilization, is the output. 
We find that you can indeed learn and capture the relationship of router configuration and both DCTCP's and Cubic's buffer sharing behavior.
To do this we use a Deep Regression model and find that our model has high accuracy ($r^2$ regression score of 0.95) with only 5\% of training data.
This means that our model is not only accurate in predicting buffer sharing behavior but is also generalizable to unseen settings since it learns with little training data.
We propose that this model can be used in future work to select optimal buffering settings for DCTCP and Cubic traffic sharing in data center buffers.

\section{Background on DCTCP}
DCTCP is a congestion control introduced roughly a decade ago for use in the data center~\cite{alizadeh2010data}.
DCTCP is designed to minimize queue occupancy by providing the congestion control with congestion notifications (ECN).

Loss signals are a delayed congestion signal because a loss signifies that router buffers are already full.
However, DCTCP uses ECN signals in undropped packets to realize that congestion is happening in the network.
Under this algorithm, routers enable ECN marking, and packets above a router buffer threshold are flagged with the ECN signal.
Once the receiver gets the ECN packet, it relays this information to the sender so that the sender is aware of network congestion.
DCTCP senders can then calculate the fraction of packets that have ECN marks and adjust sending rates accordingly.
In this way, ECN is a faster signal than loss in the network, and router buffer utilization and losses are kept much lower than with loss-based congestions.

\section{Motivation and Related Work}

First, DCTCP is used in modern data center networks and much work has been performed to analyze the performance of DCTCP.
However, data centers also contain traffic from other TCP congestion algorithms in their networks~\cite{yan2021acc}.
As motivation for this work, we highlight two main reasons and challenges for measurement of DCTCP and Cubic performance under diverse router configurations in the data center: lack of research on DCTCP and Cubic performance when sharing buffers and the effects of RED~\cite{floydRED} and drop-tail router configurations are not immediately clear when both DCTCP and Cubic traffic compete with each other.

\subsection{Gap in Related Work on DCTCP Buffer Sharing}
TCP fairness is an important issues across many different types of networks.
In WAN networks, much work has been performed to understand the fairness attributes of Cubic~\cite{abdeljaouad2010performance, kozu2013improving, ha2008cubic}, BBR~\cite{ware2019modeling, sasaki2018tcp, song2020bbr, zhang2018modest, nashbbr}, and other popular congestion control algorithms~\cite{abdeljaouad2010performance, sasaki2018tcp}.
For example, BBRv2~\cite{googleBBRv2alpha} has been introduced and claims to fix many TCP fairness issues seen in BBR.
However, fewer works have explored the area of DCTCP buffer sharing and interaction with other congestion algorithms. 
Below we present works that analyze DCTCP buffer sharing and how they differ from our work.

First, the authors of DCTCP claim that fairness is a non-issue since DCTCP is deployed in a controlled network (ie. the data center environment)~\cite{alizadeh2010data}.
However, data centers are often not as homogeneous as once thought and can contain many competing protocols~\cite{yan2021acc}.
While isolation of traffic in the data center seems like an easy solution, it is often not feasible because DC routers contain smaller router buffers and limited router queues.
Limited router queues are used to prioritize traffic classes such as via QoS scheduling (i.e. dscp traffic management) making it difficult to both prioritize traffic and separate different TCP algorithms.
Secondly, since data center routers have smaller buffers to begin with, router buffer space becomes a space commodity that is 1) needed by Cubic to minimize losses and ramp up sending rate and 2) used by DCTCP to absorb short, quick bursts of traffic.

As a result several works look at the coexistence of DCTCP and other WAN protocols~\cite{ganji2020characterizing, zhang2019ddt, zhang2018designing, kühlewind2014using, irteza2014Coexistence}. 
\cite{irteza2014Coexistence} looks at coexistence between DCTCP and Reno congestion controls where both algorithms react to ECN. 
Another work~\cite{kühlewind2014using} looks at DCTCP and Cubic coexistence in WAN settings.
Other works~\cite{zhang2018designing, zhang2019ddt} design their own TCP variants and AQMs to improve coexistence of ECN and non-ECN data center traffic.
One main shortcoming with these works is that they do not consider the dual-RTT nature of congestion in data centers where DCTCP is used for short-RTT flows and Cubic is used for longer-RTT flows~\cite{ghabashnehIMC22}.
This further complicates coexistence because many TCP congestions are known to have RTT-unfairness against themselves. 
In fact, ~\cite{ghabashnehIMC22} measures dual-RTT traffic in a data center, but only looks at bursty traffic in this context.  

Some works dynamically tune ECN router configurations in the data center, but do not explicitly consider TCP fairness and only set utilization targets~\cite{yan2021acc}.
Other works only analyze fairness of DCTCP against itself but does not consider inter-protocol performance of DCTCP~\cite{alizadeh2011analysis, tsiknas2021fairness}.

Therefore, we find a gap analysis of DCTCP and Cubic performance when buffer sharing in the data center.

\subsection{Optimal Router Configurations are Unclear}
Another challenge motivating this work is that the operating mechanisms of both Cubic and DCTCP under a shared router buffer are at odds with each other.
Data center traffic workloads often require a combination of both low latency but also high throughput.
In the presence of Cubic and DCTCP algorithms, these two goals are are conflicting as we explain below:

First, by nature Cubic saturates router buffers until the Cubic traffic achieves a loss (ie. full buffer).
Cubic achieves high throughput by growing its congestion window and sending more data every RTT, ultimately saturating router buffers.
However, Cubic needs sufficiently free buffer space to achieve high throughput otherwise its throughput will be cut continually due to losses.

On the flip side, DCTCP cuts sending rates at the sign of any congestion in router buffers.
This happens because routers mark packets with ECN when the ECN buffer threshold has been surpassed.
Once packets are marked with congestion bits, senders diminish sending rates to avoid saturating router buffers.
Thus, unlike Cubic, DCTCP operates best when router buffers are empty, which makes it difficult to find a balance in buffer management settings when Cubic and DCTCP algorithms are sharing router buffers.

These differences are further exacerbated with the requirements of latency-sensitive vs. throughput-heavy flows in the data center.
Latency-sensitive flows benefit from empty buffers that minimize queuing delay in the data center.
On the other hand, the goal of throughput-heavy flows are agnostic to router buffer utilization. 

Because of the above complexities, it is difficult to adjust router configurations, like ECN and RED parameters, to please both Cubic and DCTCP traffic and to accomplish healthy buffer sharing for both algorithms.

\section{Key Router Configurations}
Since we are looking at DCTCP and Cubic traffic, this work focuses on the effect of two router settings: ECN and RED thresholds.
These two parameters are chosen since they are largely available in commodity routers and can be readily deployed without network modification.
For DCTCP, ECN threshold controls the buffer level at which the data center router begins marking packets as congested.
For Cubic, we set RED's minimum and maximum drop thresholds.
RED increases the drop probability as the buffer occupancy increases throughout the min-to-max threshold ranges.

\section{Experimental Methodology}

Our goal is to understand the performance implications when changing buffer settings on DCTCP and Cubic streaming traffic in the data center.
We use NS3 simulations in order to perform large-scale experiments while varying multiple settings.
We first present our NS3 simulation setup, including topology, streaming workload, and data collection.
Next, we present an overview of our router buffer implementation and well as run-time challenges of this simulation setup.

\begin{figure}[!t]
    \centering
    \includegraphics[width=\linewidth]{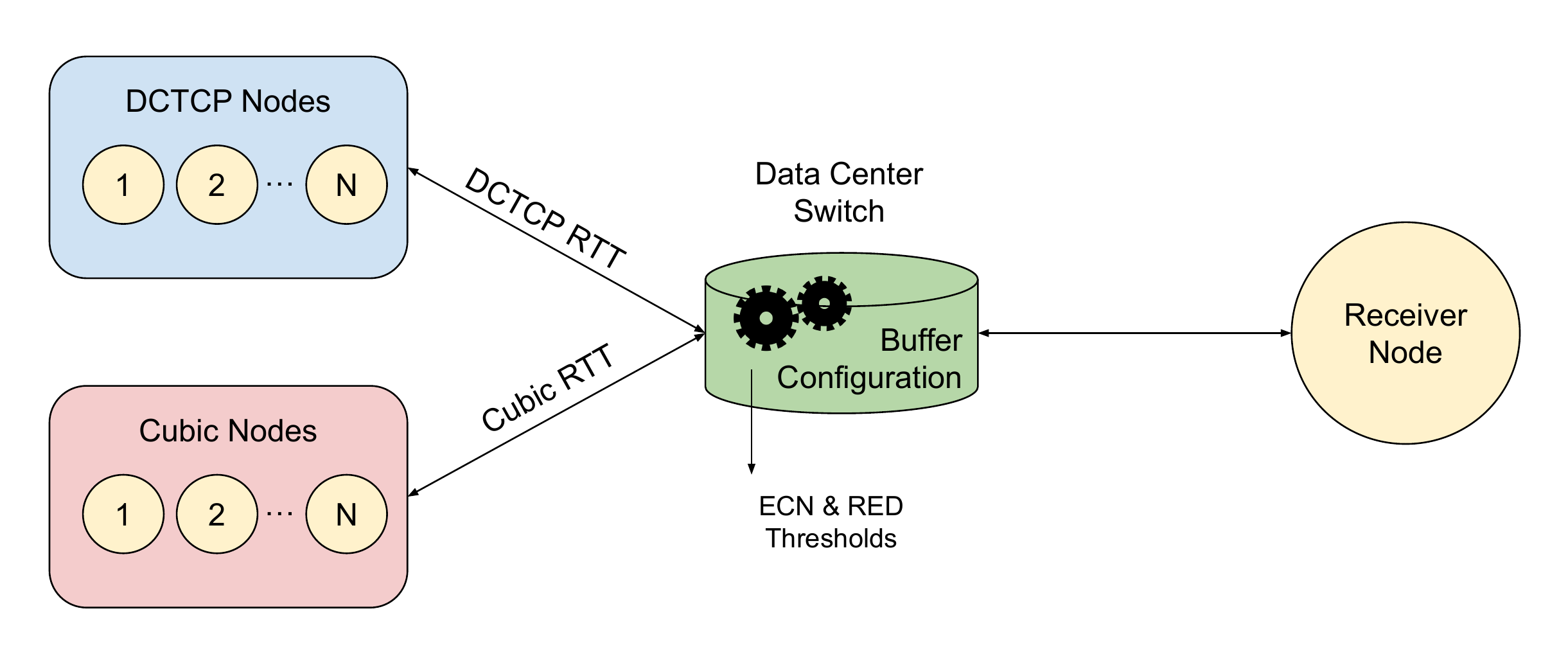}
    \caption{Topology of NS3 simulations.}
    \label{fig:dctcp_ns3_topology}
\end{figure}

\subsection{Topology}
NS3 is setup with a simple dumbbell topology as shown in Figure \ref{fig:dctcp_ns3_topology}.
In this topology, there are two groups of sender nodes, DCTCP and Cubic, which are each configured with their respective congestion controls.
Each DCTCP and Cubic node is connected to the receiver node via the switch. 
Individual links between sender nodes and the receiver nodes are configurable in terms of link rate (bandwidth) and delay (rtt).
The link between the receiver node and switch is configured with a 50us delay similar to the delay between a data center server and its parent switch.

\subsection{Workload}
The number of flows per sender is obtained as input to the simulation.
Every sender node creates several NS3 BulkSendApplications, which are iPerf-like streaming flows that are configured to send data to the receiver node.
These streaming applications continually send as much data possible until the simulation is ended.
We set large TCP send and receive buffers on sender and receiver nodes so that flows are not TCP-buffer limited.
Flows also randomly started on sender nodes to avoid synchronization artifacts.

\subsection{Router Buffer Implementation}
We find that NS3 does not contain buffering models for shared DCTCP and Cubic queues. 
To experiment with DCTCP and Cubic buffer sharing, we implement our own buffer as an NS3 queue disc class using NS3's traffic control layer.
Our queue disc employs both ECN and RED behavior for DCTCP and Cubic traffic respectively.
When packets are enqueued, the ECT bit is checked to determine if the packet is a DCTCP or Cubic packet. 
If ECT is set then packets are marked with ECN if above the ECN threshold.
Otherwise, packet are Cubic packets and are dropped according to RED~\cite{floydRED} settings: min threshold, max threshold, and drop probability.
All packets from sender nodes pass through a single queue before being sent to the receiver node.

\subsection{Data Collection}
We employ data collection at three different layers: application, TCP, and router. 
At the application, we record total goodput and runtime from each individual sender and receiver. 
From TCP, we record packets sent, acked, and dropped, and retransmitted.
From the router queuing layer, we collect the number of packets enqueued, dequeued, dropped (including reason for dropping), and marked ECN.
We also record router buffer utilization.

Data is collected in an periodic manner using a poisson process. 
This is so that data is not biased towards specific time interval and is instead randomly sampled.

\subsection{Implementation and Overhead}
The above experiment setup is implemented in NS3 version 3.33 which includes a TCP Cubic model.
We find that each experiment takes on average up to 4GB of RAM and 4 hours to complete for a 2 minute simulation with 20 senders (10 DCTCP and 10 Cubic senders).
Experiment time and overhead also scales with BDP such that high bandwidth and rtt experiments take longer to run.

\begin{table}[t]
\centering
\begin{tabular}{|l|l|}
\hline
Parameter              & Values                     \\ \hline
Cubic RTTs             & 25ms, 50ms, 100ms          \\ \hline
DCTCP RTTs             & 50us                       \\ \hline
Line Rates             & 5Gbps, 12.5Gbps, 25Gbps    \\ \hline
Buffer Size            & 1.8MB                      \\ \hline
ECN Thresholds         & 0KB, 20KB, 40KB... 400KB   \\ \hline
Min/Max RED Thresholds & 0KB, 100KB, 200KB... 1.8MB \\ \hline
Drop Probability       & 5\%                        \\ \hline
\end{tabular}
\caption{Parameters ranges used in buffer sharing simulation. Intersection of all parameters is about 10K unique experiments.}
\label{tab:dctcp_params}
\end{table}

\section{Evaluation}
The goal of this work is to improve the sharing behavior of DCTCP and Cubic traffic across many network conditions by selecting optimal router configurations.

\subsection{Parameter Selection}
To do this, we perform around simulations for over 10000 unique buffer settings spanning many combinations of all possible combinations of the parameters in Table \ref{tab:dctcp_params}.
We experiment with long Cubic RTTs since Cubic is used for long-distance traffic between data centers while DCTCP is used for traffic within the data center~\cite{ghabashnehIMC22}.
Buffer size is chosen as 1.8MB since this represents production settings for commonly used Broadcom switches~\cite{ghabashnehIMC22}.
Lastly, we experiment with a wide range of viable ECN and RED thresholds to completely cover all possible router configuration combinations.
ECN threshold is limited to 400KB (instead of the maximum of 1.8MB) as we find that ECN threshold has minimal effect above this range.
We maintain one RED drop probability at 5\% so as to minimize the number of experiments and propose that future work should look at different drop probabilities.

\subsection{Experiment Setup}
Experiments are configured with 10 DCTCP and Cubic senders each and 10 flows per each sender for a total of 100 Cubic and DCTCP flows.
Each experiment runs for 2 minutes of simulation time to allows even 100ms Cubic flows sufficient RTTs to reach steady state and convergent behaviour.
Upon experiment completion, an archive file is created with experiment configuration and data.
Data collection configured to run a snapshot on average every 10ms of simulation time.

\begin{figure}[!t]
    \centering
    \includegraphics[width=\linewidth]{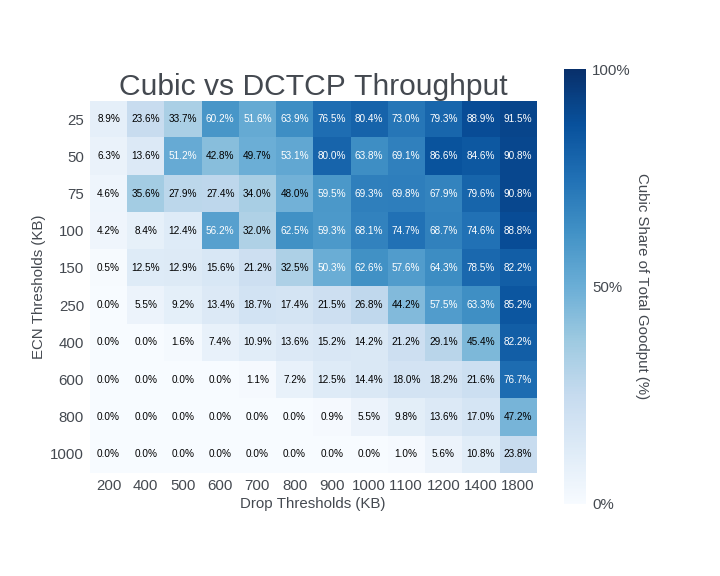}
    \caption{Percentage of Cubic throughput across range of ECN and Drop-Tail buffer settings.}
    \label{fig:dctcp_tput}
\end{figure}

\subsection{Performance under Drop-Tail Settings}
In this section, we present examples of DCTCP and Cubic performance while under drop-tail settings where Min and Max RED thresholds are equal.
We also include ECN thresholds above 400KB since to show how DCTCP largely dominates the buffer with these settings.

\begin{figure}[!t]
    \centering
    \includegraphics[width=\linewidth]{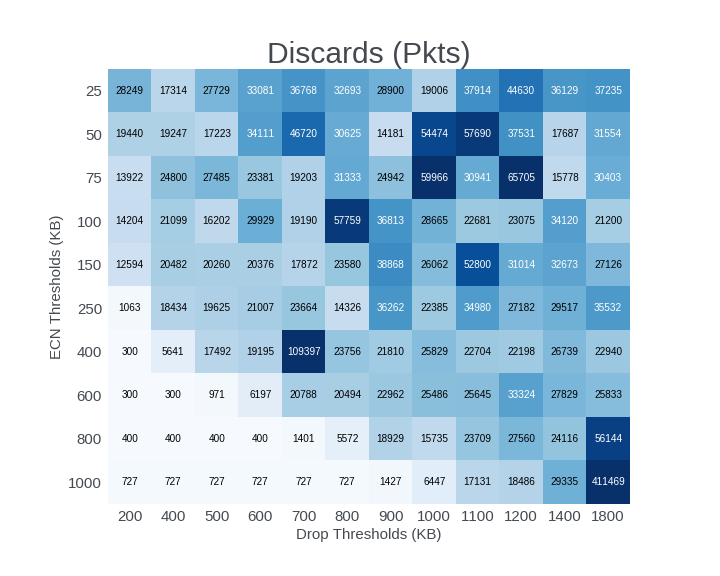}
    \caption{Packet drops across range of ECN and Drop-Tail buffer settings.}
    \label{fig:dctcp_discards}
\end{figure}

Figures \ref{fig:dctcp_tput} and \ref{fig:dctcp_discards} show the share of Cubic throughput and total packet drops with a wide range of ECN and drop-tail settings. While we omit RED results, we explain differences between RED and drop-tail in the next subsection.

We make several observations about Cubic and DCTCP traffic and buffer settings.
First, lower ECN thresholds increase DCTCP throughput. 
Figure \ref{fig:dctcp_tput} shows that smaller ECN thresholds (top of the figure) have higher DCTCP throughput shares, and larger ECN thresholds have lower DCTCP throughput shares.
Similarly, and a higher drop-tail threshold increases Cubic throughput. 
Figure \ref{fig:dctcp_tput} shows that the right side (larger Drop thresholds) favor Cubic share of throughput, and lower drop thresholds (left side) reduce Cubic's throughput share.
Another observation is that the trade off for increased Cubic throughput is an increase in packet drops.
Figures \ref{fig:dctcp_tput} and \ref{fig:dctcp_discards} show a correlation between increase in Cubic's throughput with packet drops (top right has higher Cubic throughput but more packet drops).

\subsection{Performance differences with RED Settings}
We also make several observations in differences between RED (i.e., when Min and Max RED thresholds are not equal) and drop-tail (i.e., equal Min and Max RED thresholds) settings.
First, RED has slightly lower packet drops than drop-tail settings. This is because Cubic flows receive an earlier loss signal than they would have with purely drop-tail settings.
Secondly, RED also has lower average buffer utilization while still maintaining similar max buffer utilization than drop-tail configurations.
This is because Cubic traffic begins receiving throttling signals near the Min RED threshold and averages remain lower than the Max RED threshold.
Therefore, RED is still effective for Cubic in the face of DCTCP competition.

Overall, it is not immediately clear if there is one optimal ECN and RED setting for DCTCP and Cubic traffic every settings produces 

\section{Modeling Buffer Settings using ML}
In this section, we first motivate the use of Machine Learning to model the relationship between buffer configurations and DCTCP/Cubic performance in the data center.
Next, we describe how we map our simulation experiments to a regression problem, including our features, training data, and regression score.

\subsection{Motivating the use of ML}
While the previous section showed a subset results for 50ms Cubic RTTs and only for drop-tail settings, the relationship between buffer parameters and DCTCP/Cubic performance is not immediately clear when varying RTTs, changing the line rate (i.e., bandwidth), and introducing RED (i.e., distinct MIN and MAX thresholds).
Two main issues that arise when increasing the parameter space of experiments are unclear effects of different buffer settings under diverse network conditions and a substantial increase in experimentation time. 
First, manual exploration of simulation results showed more complex relationships between buffer settings and resulting TCP performance making it difficult to decide on specific buffer settings.
Secondly, the experiment space increases substantially with every extra parameter that is added to the experiment exploration space.
Since simulation runs take on around 4 hours and large amounts of RAM, it becomes costly to run these experiments across large amounts of parameters.
Therefore, a machine learning model would be able to learn the relationships between buffer settings, network conditions, and DCTCP/Cubic performance while providing quick inference.
This would allow a much more rapid exploration of the experiment search space.

\subsection{ML Problem Formulation and Results}
\subsubsection{Input Features}
Our input features to the ML problem includes router configurations and network conditions.
We include router configurations as these are parameters that can be fined-tuned by network operators to improve DCTCP and Cubic's performance.
For router configurations, we consider the ECN threshold, Min RED threshold, and Max RED threshold.
While network conditions are static, they also affect the buffer sharing behavior of DCTCP and Cubic.
Therefore we include the Cubic RTT and Line Rates (bandwidth) as inputs.

\subsubsection{Problem Definition}
We formally define the ML problem of learning the relationship between buffer settings, network conditions, and DCTCP/Cubic shared buffer performance as a regression problem.
These above input values are mapped via regression to the following output values: 1) Cubic's share of throughput 2) Packets Dropped 3) Average Buffer Utilization 3) Max Buffer Utilization.
Cubic's share of throughput is chosen since this value is a proxy for DCTCP and Cubic performance where short-RTT DCTCP is able to quickly react to Cubic's under-utilization of bandwidth and fill the remaining unused bandwidth.
Alternatively, future learning work can consider learning fairness metrics like Min-max fairness or Jain's fairness index.
Packets dropped is predicted in order to understand the contention between DCTCP and Cubic when sharing data center buffers.
Lastly, average and maximum buffer utilization are of interest because ECN and RED thresholds largely limit DCTCP and Cubic from utilizing excess buffers.
Average utilization alone does not tell the complete buffering picture since RED's Max threshold significantly limits Cubic's maximum buffer utilization.

\subsubsection{Generalizing the Model via Training}
The next goal of this model is to be highly generalizable yet accurate since the model will be used instead of actual simulation experiments.
Therefore, we choose to train the model with limited data and train the model with a maximum of 5\% of data.
In our experiment dataset of around 10K different settings, we train the model with around 400 experiments.
This is to make sure that our model is in fact learning the above defined problem.

\begin{table}[t]
\centering
\begin{tabular}{|l|l|l|l|}
\hline
\textbf{Model Type} & \textbf{Training Data} & \textbf{Training Split} & \textbf{$r^2$ score} \\ \hline
Shallow & Single Condition                       & 1\% & 0.95 \\ \hline
Shallow & Multiple Conditions (w/o Packet Drops) & 5\% & 0.75 \\ \hline
Deep    & Single Condition                       & 1\% & 0.99 \\ \hline
Deep    & Multiple Conditions                    & 5\% & 0.85  \\ \hline
Deep    & Multiple Conditions (w/o Packet Drops) & 5\% & 0.92 \\ \hline
\end{tabular}
\caption{Different ML learning approaches used to model relationship between router configurations and buffer sharing performance. While Shallow models can capture interaction for single network conditions, Deep Learning is best when analyzing a range of network conditions.}
\label{tab:dctcp_mlapproaches}
\end{table}

\subsubsection{Model Results}
We present a variety of ML learning approaches in Table~\ref{tab:dctcp_mlapproaches} and explain the reasoning for choosing our final model.
Since this problem is a framed as a regression, we evaluate models with $r^2$ score, mean-squared error(MSE), and mean absolute error(MAE). 
While, Table~\ref{tab:dctcp_mlapproaches} provides only $r^2$ scores, MSE and MAE results followed similar trends. 

First, we find that shallow models are sufficient to capture the relationship between buffer settings and both Cubic throughput and buffer utilization with high accuracy ($r^2$ score of $>0.95$) when trained with even as little as 1\% of data.
However, when adding data with different network conditions with the same goal of high accuracy (e.g., $r^2 > 0.9$) we find that shallow models provide substantially lower accuracy (e.g., $r^2 <= 0.75$) even with 5\% of training data.
We first attempt to remove packet drops as an output because with RED, packet drops occur randomly and may increase the overall noise in output data.
We observer that while removing drops increases the overall model score, overall accuracy is still relatively low.

As a result, we use resort to a two-layer Deep MLPRegressor model with the lbfgs optimizer~\cite{liu1989limited} to learn under diverse network conditions.
This model provides average $r^2$ scores of $0.92$ when trained with 5\% of experiment data.
Further, we find that root mean squared errors and mean absolute errors for buffer utilization are smaller than 100KB with this deep model, showing that it can accurately predict buffer utilization.
Even when including packet drops as an output, this deep regressor model can predict with an average $r^2$ score of 0.85.

\section{Future Direction using our ML Model}
Since we demonstrate that our ML Model captures the relationship between buffer settings and network conditions as they relation to DCTCP/Cubic buffering performance, we now present future directions for using this model.
The main benefit of our model is that is takes small amounts of training data to infer accurate results about buffering performance.
As a result, our model can be used to learn much larger search spaces including wider ranges of RTTs, bandwidth, and buffer setting combinations.
This will largely simplify the time and cost overhead of running simulations to capture these results.

Our model can also be used by network operators to find optimal buffer settings for their data center buffers.
Since our model provides fast inference, this model can be used in conjunction with an objective or cost function to explore buffer setting search space looking for optimal buffer settings.

\section{Conclusion}

In this chapter, we explore data center buffer sharing performance of both DCTCP and Cubic congestion control algorithms
We first run large scale simulations to understand the buffering performance of both algorithms under a diverse set of ECN and RED settings and under different network conditions.
Afterwards, we capture the relationship between ECN and RED settings on DCTCP and Cubic performance using Deep learning.
We formulate the problem as regression where the inputs are ECN thresholds, RED thresholds, link rates, and RTT; and the output is Cubic throughput, average/max buffer utilization, and loss rates.
We find that our ML model is able to learn with only a fraction (5\%) of data while maintaining high accuracy in regression ($r^2$ of 0.95).
Finally, we suggest that this model can be used to further tune buffer settings in the data center.


\bibliographystyle{unsrt}  
\bibliography{references}

\end{document}